\documentclass[onecolumn]{aastex631}    
\usepackage{threeparttable}
\usepackage{natbib}
\usepackage{amssymb,amsmath}
\usepackage{latexsym}
\usepackage{verbatim}
\usepackage{mathtools}
\usepackage{multirow}
\usepackage{empheq}
\usepackage{bm}
\usepackage{color}
\usepackage{CJK}
\usepackage{longtable}
\usepackage{hyperref}
\usepackage{comment}

\newcommand{\Ha}{H$\alpha$\,$\lambda$6563}

\newcommand{\OIIIopt}{[O\,{\sc iii}]\,$\lambda$5007}
\newcommand{\NIIopt}{[N\,{\sc ii}]\,$\lambda$6583}
\newcommand{\OIopt}{[O\,{\sc i}]\,$\lambda$6300}
\newcommand{\SII}{[S\,{\sc ii}]\,$\lambda$6724}

\newcommand{\LCII}{$L_{\rm [C\,{\scriptsize \textsc{ii}}]}$}
\newcommand{\LOI}{$L_{\rm [O\,{\scriptsize \textsc{i}}]63\,\mu{\mathrm m}}$}
\newcommand{\LFIR}{$L_{\rm FIR}$}

\newcommand{\NI}{[N\,{\sc ii}]$\ 122\ \mu \rm m$}

\newcommand{\CII}{[C\,{\sc ii}]$\ 158\ \mu \rm m$}
\newcommand{\OI}{[O\,{\sc i}]$\ 63\ \mu \rm m$}

\newcommand{\C}{[C\,{\sc ii}]}
\newcommand{\N}{[N\,{\sc ii}]}
\newcommand{\OIII}{[O\,{\sc iii}]}

\newcommand{\kms}{km\,s$^{-1}$}
\newcommand{\mum}{$\mu$m}

\shorttitle{ISM Properties in NGC~1222}  
\shortauthors{Liu et al.}            
\begin{document}
\title{Properties of Interstellar Medium in the S0 Galaxy NGC 1222: Evidence for Shocked-enhanced Line Emission\footnote{Released on XX, XX, XXXX}}

\correspondingauthor{Yinghe Zhao}
\email{zhaoyinghe@ynao.ac.cn}

\author{Jiamin Liu}
\affiliation{Yunnan Observatories, Chinese Academy of Sciences, Kunming 650216, China}
\affiliation{University of Chinese Academy of Sciences, Beijing 100049, China} 

\author{Yinghe Zhao}
\affiliation{Yunnan Observatories, Chinese Academy of Sciences, Kunming 650216, China}

\affiliation{Key Laboratory of Radio Astronomy and Technology (Chinese Academy of Sciences), A20 Datun Road, Chaoyang District, Beijing, 100101, P. R. China} 

\affiliation{Key Laboratory for the Structure and Evolution of Celestial Objects, Chinese Academy of Sciences, Kunming 650216, China}

\author{Kai-Xing Lu}
\affiliation{Yunnan Observatories, Chinese Academy of Sciences, Kunming 650216, China}

\affiliation{Key Laboratory for the Structure and Evolution of Celestial Objects, Chinese Academy of Sciences, Kunming 650216, China}

\author{Jin-Ming Bai}
\affiliation{Yunnan Observatories, Chinese Academy of Sciences, Kunming 650216, China}

\affiliation{Key Laboratory for the Structure and Evolution of Celestial Objects, Chinese Academy of Sciences, Kunming 650216, China}

\begin{abstract}
In this paper we present a comprehensive study on the properties of the interstellar medium in NGC~1222, a star-forming early-type merging galaxy that forms a triple system, using optical and far-infrared (FIR) spectroscopic, and multiband photometric data. The fit to the spectral energy distribution reveals a high dust content in the galaxy, with a dust-to-stellar mass ratio of $M_\mathrm{dust}/M_\star\sim3.3\times10^{-3}$ that is 40$-$90 larger than the mean value of local S0 galaxies. By comparing the observed optical emission line ratios to shock models, we suggest that a merger-induced shock, which is further supported by the higher-than-average \OI- and \CII-to-PAH ratios, plays a role in heating the gas in NGC~1222. We also show evidence for gas inflow by analysing the kinematic properties of NGC~1222. 
\end{abstract}


\section{Introduction}          
\label{sect:intro}
Local early-type galaxies (ETGs; i.e. elliptical and S0) generally have ``red and dead" stellar populations since they contain little cold gas to fuel star formation (SF). However, $20\%$ of the 48 SAURON ETGs (\citealt{deZeeuw+etal+2002}) show evidence of recent SF and gas reservoirs (\citealt{Combes+etal+2007}; \citealt{Shapiro+etal+2010}). More recently, \citet{Young+etal+2014} have shown that $\sim$20\% and $\sim$30\% of ETGs in the ATLAS3D (\citealt{Cappellari+etal+2011}) sample (260 sources) contain molecular and atomic gas, respectively, with the largest gas-to-stellar mass ratio ($M_{\mathrm {gas}}/M_\star$) reaching $\gtrsim$0.1. 

Nevertheless, \cite{Davis+etal+2014} find that molecular gas-rich ETGs have systematically lower SF efficiency (SFE) than normal spiral galaxies by a factor of $\sim$2.5. A number of different mechanisms, such as suppressing gas inflow and cooling (e.g., shock heating: \citealt{Birnboim2003, Lanz+etal+2016}; feedback by active galactic nuclei (AGNs): \citealt{Croton+etal+2006}), removing/heating cold gas (\citealt{Di+etal+2005, Hopkins+etal+2006}), and stabilizing gas reservoirs (\citealt{Martig+etal+2009, Fang+etal+2013, Tacchella+etal+2015}), have been postulated for quenching SF in local ETGs. 

Indeed, studies on individual objects have revealed the need for diverse mechanisms responsible for the SF suppression in ETGs. For example, \cite{Alatalo+etal+2015} suggest that the AGN-driven bulk outflow through a process of turbulent injection could account for the extreme SF suppression, by a factor of $\sim50-150$ compared to normal star-forming galaxies, in the ETG NGC~1266. For another gas-rich ETG, NGC~3665, our previous study (\citealt{Xiao+etal+2018}) show that the SF suppression (by a factor of $\sim$10) is most possibly caused by its compact, massive bulge via stabilizing cold gas (i.e., ``morphology quenching"). 

To further explore possible mechanism(s) for the SF suppression in ETGs, in this paper we present multi-band studies on the properties of the interstellar medium (ISM) in NGC~1222, one of the four ETGs (i.e., NGC~1222, NGC~1266, NGC~2764 and NGC~3665) having the largest molecular gas ($\log M_{\rm H_2} (M_\odot) = 9.33$; \citealt{Alatalo+etal+2013}) reservoirs in the ATLAS3D sample. The ISM plays a key role in galaxy formation and evolution. The neutral ISM provides the raw material of SF, and massive stars in turn can enrich the heavy elements and heat up the ISM through the photoelectric process (PE; \citealt{Tielens+Hollenbach+1985}), influencing the next rounds of SF. With the advent of the {\it Herschel Space Observatory} (hereafter {\it Herschel}; \citealt{Pilbratt+etal+2010}), it allows us to investigate excitation tracers in the far-infrared (FIR), such as the \CII\ and \OI\ fine-structure emission lines, which can provide critical diagnostic tools for the physical conditions of the star-forming ISM (e.g., \citealt{Stacey+etal+1991,Kaufman+etal+1999,Malhotra+etal+2001}). In addition, \cite{Lu+etal+2014} find that the FIR line-to-continuum ratios (also see \citealt{Appleton+etal+2013,Alatalo+etal+2014,Fadda+etal+2023} for other individual examples) can be a good indicator of shocks unrelated to the current SF, such as those associated with radio jets, AGN-driven gas outflows, or galaxy–galaxy collision.

\begin{figure}[tb]
\centering
 \includegraphics[width=0.47\textwidth,bb=10 109 525 655]{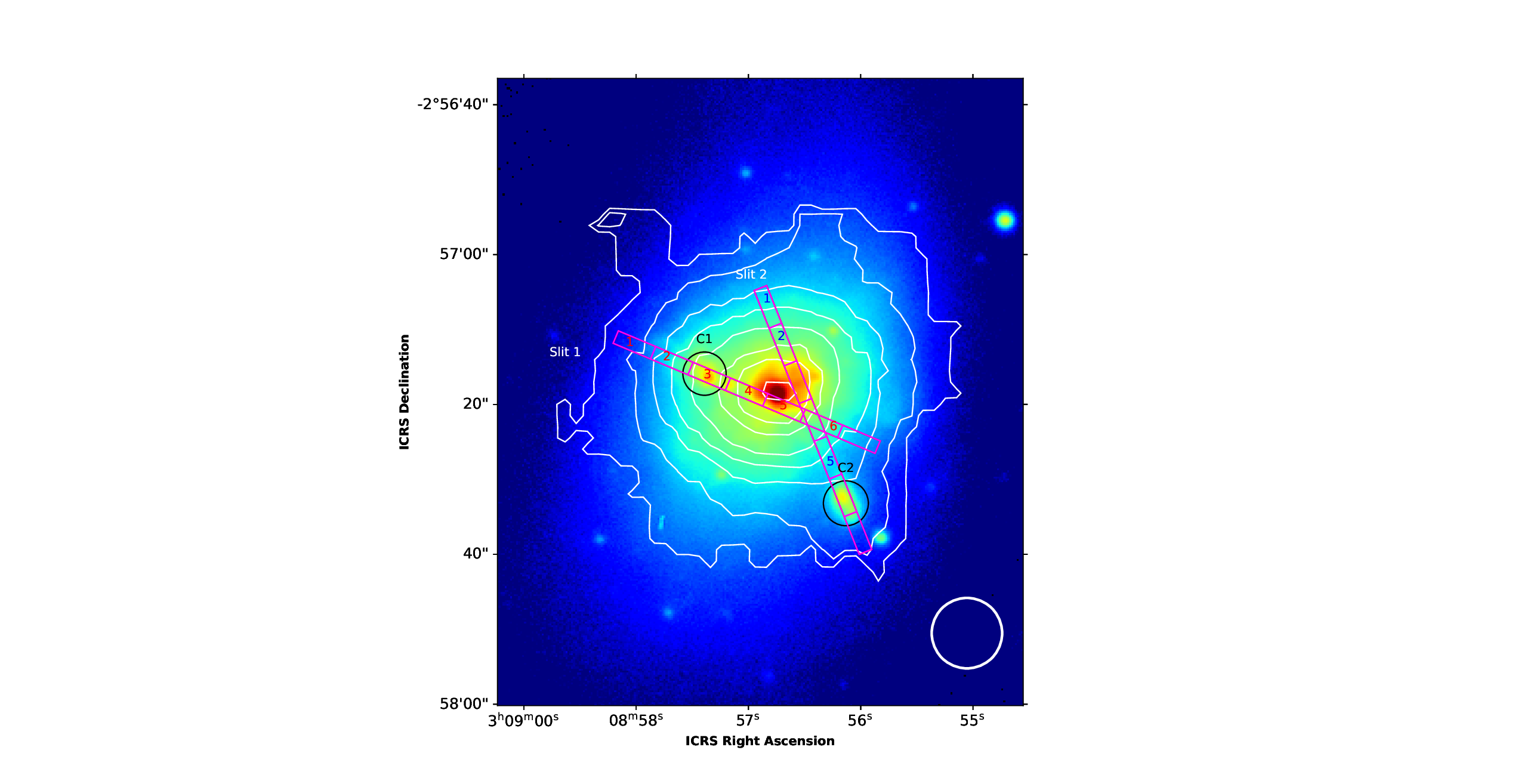}
    \caption{$r$-band image from the DESI survey, overlaid with our two long slit positions (magenta rectangles), which are set to cover a bright reference star and one knot (C1/C2) each, and the PACS \OI\ intensity contours ([0.01, 0.055, 0.15, 0.3, 0.6, 1.2, 1.6, 2.1]$\times10^{-17}\,{\mathrm{W}}\,{\mathrm{m}}^{-2}$; white lines). The boxes ($5\arcsec.4 \times 1\arcsec.8$) show the apertures we extract 1d spectra. C1 is located in aperture 3 of slit 1 and C2 is located in aperture 6 of slit 2. The white circle (bottom right) gives the beam size (${\mathrm{FWHM}}\sim 9\arcsec.4$) of PACS for the \OI\ map.}
    \label{r_band_image}
\end{figure}

NGC~1222 is one of the two objects (NGC~1222 and NGC~2764) that are also rich in atomic gas ($\log M_{\rm H\,\scriptsize{\textsc {i}}}\,(M_\odot)=9.46$; \citealt{Young+etal+2018}). As a slowly rotating ETG (\citealt{Emsellem+etal+2011}), it is the only one detected in both cold atomic and molecular gas emission in \cite{Young+etal+2018}. \cite{Petrosian+Burenkov+1993} identify NGC~1222 as a merging system, comprising a starburst nucleus of high surface brightness and two dwarf nuclei C1 and C2, as shown by the $r$-band image (Figure \ref{r_band_image}) from the DESI Legacy Imaging Surveys (\citealt{Dey+etal+2019}). From Figure \ref{r_band_image} we can clearly see stellar shells and tidal tails (also see \citealt{Duc+etal+2015}). Combining with its disrupted stellar and gaseous kinematics, the molecular gas in NGC~1222 is likely to be acquired via external accretion (\citealt{Alatalo+etal+2013}). Therefore, NGC~1222 is a good system to investigate the influence of merger on the interstellar medium (ISM) and SF properties within individual galaxies (e.g., \citealt{Rich+etal+2011,Appleton+etal+2013,Alatalo+etal+2014,Fadda+etal+2023, Zhang+etal+2024}). 

Unlike NGC~1266 (\citealt{Alatalo+etal+2015}) and NGC~3665 (e.g., \citealt{Xiao+etal+2018}) showing a severely suppressed SF, however, NGC~1222 is classified as a starburst system (\citealt{Balzano+etal+1983}). As presented in \cite{Davis+etal+2014}, it has a SF rate (SFR) of $\sim$3-6~$M_\odot\,{\mathrm {yr^{-1}}}$ (depending on the adopted SFR calibrator) and SFR surface density of $\Sigma_{\mathrm{SFR}}\sim2.0-3.5$~$M_\odot$~yr$^{-1}$~kpc$^{-2}$. Despite relatively high SFR, the SFE of NGC~1222 is still lower than normal spiral centers by a factor of $>$3.5 given its abundant gas content (we note that only the H$_2$ mass has been taken into account when calculating the gas surface density in \citealt{Davis+etal+2014}). Hence, the SF activity is also suppressed in NGC~1222.

In this paper we conduct a comprehensive study on NGC~1222, using muti-band photometric data to fit the spectral energy distribution (SED), and exploiting spectroscopic data in the optical (observed by the Li Jiang 2.4m telescope), mid-IR (obtained by the the {\it Spitzer} Space Telescope; \citealt{Werner+etal+2004}) and FIR (observed by {\it Herschel}) to perform different diagnostic diagrams. We explore the power source of emission lines, and investigate kinematic properties of the ISM in NGC~1222. Our paper is organized as follows: we describe the observations and data reduction in section~\ref{sect:Obs}, present our results and analysis in section~\ref{sect:results}, discuss possible contributions to the line excitation in section~\ref{sect:discuss}, and summarize our main findings in the last section.

\section{Observations and Data Reduction}
\label{sect:Obs}
\subsection{Optical Spectroscopy}
\label{Opt-intro}
Long-slit optical spectroscopy was conducted with the Yunnan Faint Object Spectrograph and Camera (YFOSC), mounted on the Lijiang 2.4-meter Telescope (LJT) at Yunnan Astronomical Observatories, Chinese Academy of Sciences, on November 29, 2018. It is equipped with a $2048\times4608$ pixel CCD with a pixel scale of $0\arcsec.283$, corresponding to a field of view of $\sim10^\prime\times22^\prime$. Observations were carried out for two different position angles (see Figure \ref{r_band_image}), which are set to cover a nearby reference star and one knot (C1/C2) for each observation. The reference star is used to investigate the observational conditions (e.g., seeing, geometric distortion), please see \citet{Lu+etal+2016} for more detail of this observation strategy. We used Grism 3 with a fixed silt width of $1\arcsec.8$, which covers the wavelength range of $3400-9100$ \AA\ with a dispersion of $\sim$2.9~\AA~pixel$^{-1}$. Standard neon and helium lamps were used for wavelength calibration and HR~718 was observed for flux calibration. 

The two-dimensional spectroscopic images were reduced using standard {\it IRAF} tools before absolute flux calibration, including bias subtraction, flat-field correction, and wavelength calibration. The accuracy of wavelength calibration and spectral resolution are checked by fitting the sky 5577 \AA\ line, which gives an accuracy of $<$1 \AA\ and a full width at half maximum (FWHM) of $\sim$16 \AA\ (i.e., a velocity resolution of 730~\kms\ at \Ha). The seeing is $\sim$3\arcsec, obtained by investigating the spectroscopic images of the reference stars. All the spectra were extracted using a uniform aperture of 19 pixels (5\arcsec.4), and the background was determined from regions on either side of the locus of NGC~1222. 

\begin{figure*}[t]
\centering
 \includegraphics[width=0.98\textwidth,bb=82 361 726 761]{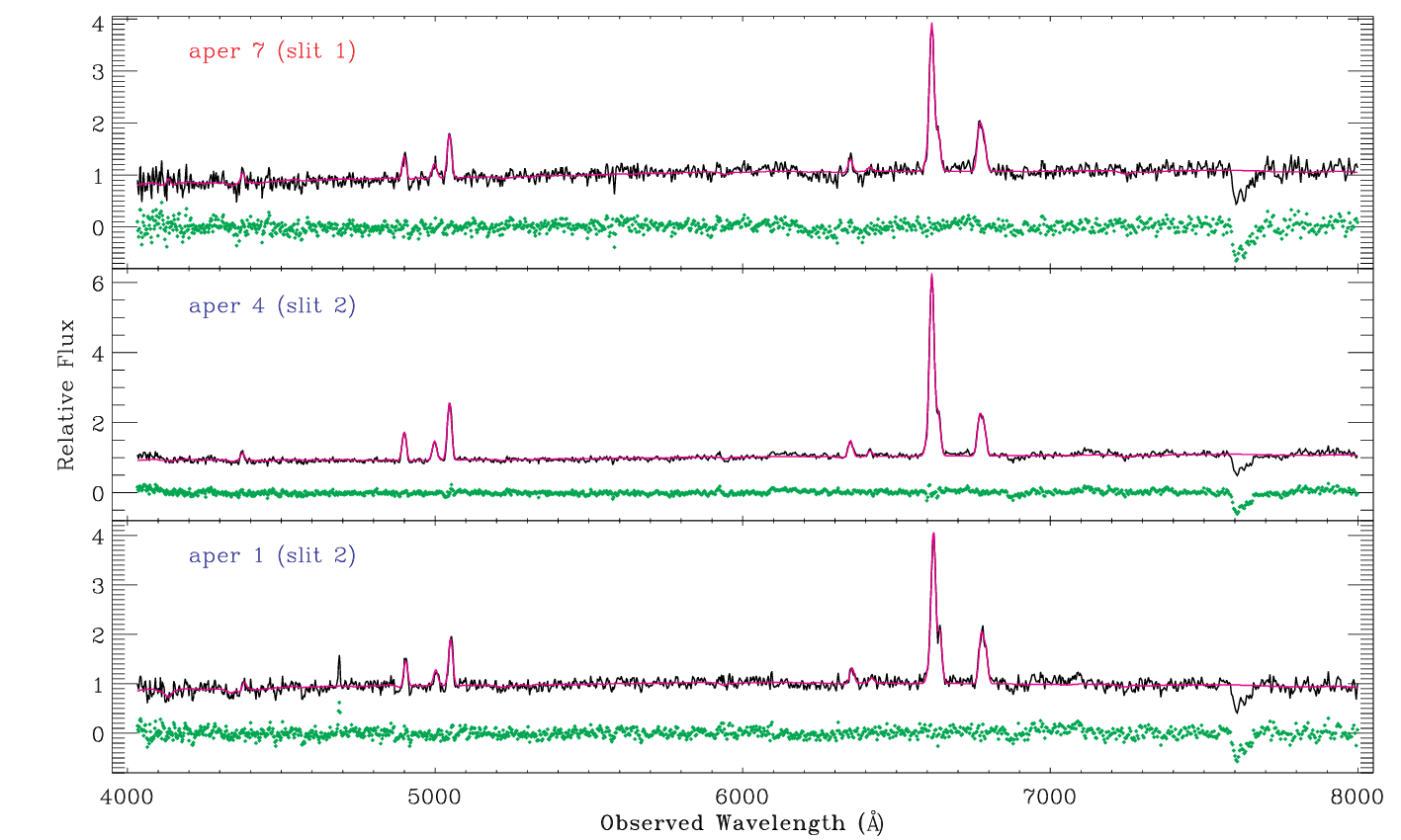}
    \caption{Observed optical spectra (black) for three apertures overlaid with the best-fitted results (stellar+emission lines; magenta) from \textsc{pPXF}. The residual spectrum is potted with green points.}
    \label{opt_spec}
\end{figure*}

To obtain the emission line fluxes, we modelled the stellar continuum with templates following the work of the SEAGal Group (\citealt{CidFernandes+etal+2005}), and simultaneously measured the gas emission using \textsc{pPXF} (\citealt{Cappellari+Emsellem+2004}). The templates are made up of 150 single stellar populations, including 25 ages (from 1 Myr to 18 Gyr) and 6 metallicities (0.005, 0.02, 0.2, 0.4, 1.0 and 2.5 $Z_\odot$), which are generated using the popular synthesis code of \citet{Bruzual+Charlot+etal+2003} with the \citet{Salpeter+1995} initial mass function, Padova 1994 models, and the STELIB library (\citealt{LeBorgne+etal+2003}). In Figure \ref{opt_spec} we show three examples of the observed (black) spectrum, best-fitting stellar (red) and line (magnenta) emissions.

\subsection{Mid- and Far-infrared Spectroscopic Data}
\subsubsection{Spitzer IRS Spectroscopy}
The mid-infrared (MIR; $5.2-38$ \mum) spectrum of NGC~1222 was observed (AOR key: 9071872) with the low-resolution mode of the Infrared Spectrograph (IRS) (\citealt{Houck+etal+2004}) on board the {\it Spitzer Space Telescope} (\citealt{Werner+etal+2004}). The processed data were downloaded from the Infrared Database of Extragalactic Observables from Spitzer (IDEOS; \citealt{Spoon+etal+2022}){\footnote{\url{http://ideos.astro.cornell.edu/}}}. As in \citet{Brandl+etal+2006}, we scaled the SL2, SL1, and LL2 spectra to match the flux density of LL1 by multiplying the stitching factors of (1.665, 1.698, 1) given in \citet{Spoon+etal+2022}. 

\begin{figure*}[t]
\centering
\includegraphics[width=0.7\textwidth,bb=-100 70 700 676]{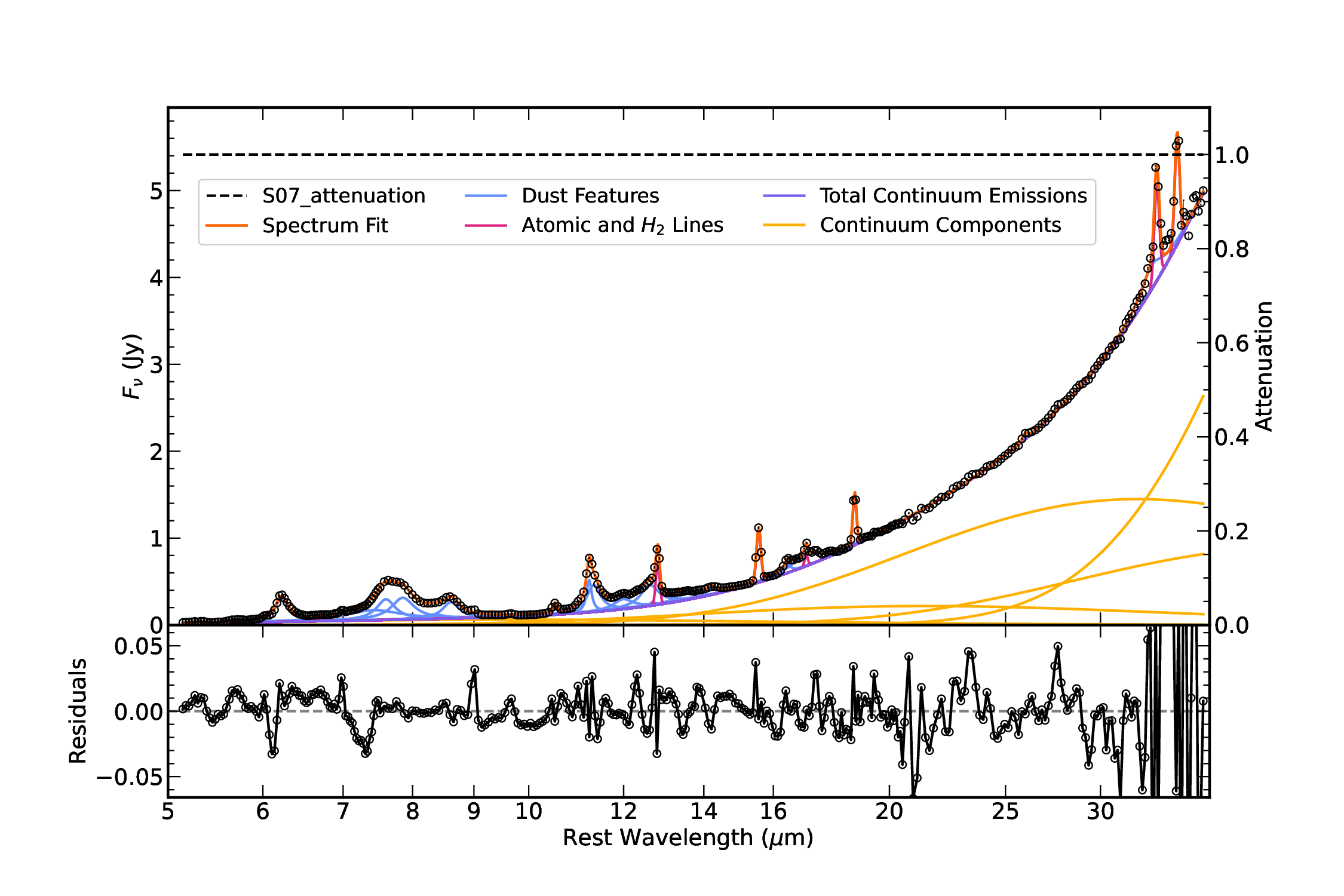}
\caption{The full decomposition result obtained from {\sc{pahfit}} for NGC~1222. The bottom panel shows the residuals obtained by subtracting the fitted model from the observed spectrum.}\label{pah_spec}
\end{figure*}

To obtain the PAH fluxes, we exploited {\sc {pahfit}} (Python version v2.1; \citealt{Lai+etal+2020}), which is a robust decomposition model first employed for extracting the integrated strengths of dust and gas features from the low-resolution IRS spectra in the MIR regime (\citealt{Smith+etal+2007}). The rest-frame spectrum was fitted with {\sc {pahfit}}, as shown in Figure \ref{pah_spec}, using the default pack file for extragalactic sources. From Figure \ref{pah_spec} we can see that the observed spectrum is well fitted and generally the residuals are small. The fluxes of the 6.2 \mum, 7.7 \mum\ Complex, 11.3 \mum\ Complex, and 17 \mum\ Complex PAH features are $6.32\times10^{-15}$, $2.37\times10^{-14}$, $6.01\times10^{-15}$, and $2.82\times10^{-15}$~W~m$^{-2}$, respectively. The integrated strength of these PAH emission features is $5.26\times10^{-14}$~W~m$^{-2}$. 

\subsubsection{Herschel PACS Spectroscopy}
\label{PACS intro}
The \OI, \NI\ and \CII\ lines of NGC~1222 were observed with the {\it Herschel} PACS spectrometer (\citealt{Poglitsch+etal+2010}) under the project OT1\_lyoung\_1 (PI: L. Young; \citealt{Lapham+etal+2017}), with total observational times of 1.2 and 0.9 hr, respectively for \OI\ (ObsID: 1342239495) and \NI/\CII\ (ObsID: 1342239496). The PACS integral-field spectrometer is composed of $5\times5$ squared spatial pixels (spaxels), each with a size of $9\arcsec.4$, and thus has a field of view (FOV) of $47\arcsec\times47\arcsec$. For \OI, \NI\ and \CII, the spatial resolutions are $\sim$$9\arcsec.4$, $10\arcsec$ and $11\arcsec.5$, respectively; and the effective spectral resolutions are about 86~\kms, 290~\kms, and 238~\kms, respectively. For NGC~1222, the data of these three lines were acquired in the mapping mode, and observations were done as $3\times3$ ($2\times2$) rasters with a step of $3\arcsec$ ($4\arcsec.5$) for the \OI\ (\NI\ and \CII) line. Therefore, the point spread function is fully sampled and we can obtain spatially resolved information on the gas in NGC~1222 utilizing these raster mapping data. 

The Level 2 data, reduced with the {\it Herschel} Interactive Processing Environment (HIPE; \citealt{Ott+2010}) version 14.2, were downloaded directly from the {\it Herschel} Science Archive{\footnote{\url{http://archives.esac.esa.int/hsa/whsa/}}}. According to PACS Products Explained Release 2.0{\footnote{\url{http://www.cosmos.esa.int/documents/12133/996891/PACS+Products+Explained}}}, we extracted the line fluxes for each spatial pixel ($1\arcsec.5$, $1\arcsec.6$ and $1\arcsec.9$ for \OI, \NI\ and \CII\ respectively) from the projected cubes by fitting the observed profile with a Gaussian function plus a linear component (for the continuum). The line is detected if its integrated flux to the fitted error ratio (S/N) is $\geq$3. The intensity contours of \OI\ are over-plotted in Figure \ref{r_band_image}. We can see that the \OI\ emission peaks in the nuclear region, but C1 and C2 also show some excess. 

To validate whether our measurements of the line fluxes are reliable, we compared our spaxel-integrated flux with those given in \citet{Lapham+etal+2017}, who obtained the line fluxes through a single spatially integrated spectrum. We found that our \OI\ and \CII\ fluxes are $\sim$4\% and 10\%, respectively, lower than the results in \citet{Lapham+etal+2017}. This might be due to the fact that a single spaxel spectrum without line detection also contributes flux to the spatially integrated spectrum. Therefore, we adopt the line fluxes in \citet{Lapham+etal+2017} when calculating the total line ratios, and use our own measurements when doing the spatially-resolved analysis. 

\begin{figure*}[tb]
 \centering
 \includegraphics[width=0.78\textwidth,bb=31 121 979 948]{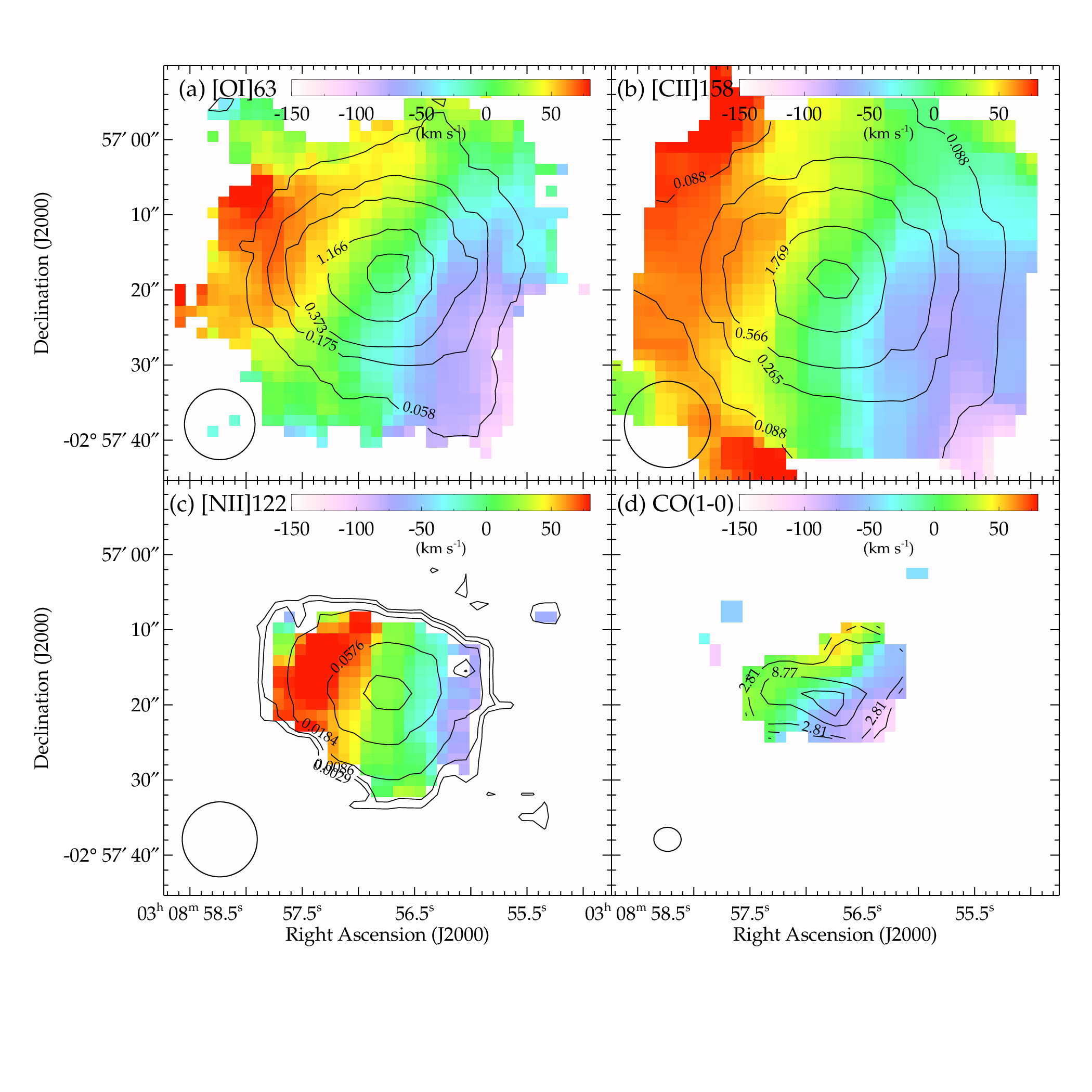}
 \caption{Line emission contours of the integrated map superimposed on the corresponding velocity field map (a systematic velocity of 2422 \kms\ has been subtracted): (a) \OI; (b) \CII; (c) \NI; and (d) CO(1$-$0). The contour levels are $[0.025, 0.075, 0.16, 0.5, 0.84]\times$~the peak flux of each map, in units of $10^{-17}$~W~m$^{-2}$ (\OI, \CII\ and \NI) and Jy~\kms~beam$^{-1}$ (CO(1$-$0)). The white circle/ellipse in the lower left corner of each panel shows the beam shape.}
 \label{pacs_lines}	
 \end{figure*}

In Figure \ref{pacs_lines}, we show the velocity field maps of these three lines for NGC~1222, with the integrated flux intensity contours overlaid. It can be seen that these velocity maps have the typical ``spider-diagram" shape, and show clear velocity gradients along the northeast-southwest direction, which characterize a rotating disk (also see the P-V diagrams in \citealt{Lapham+etal+2017}). Furthermore, the ionized gas emission (i.e., \NI\ line) has a much smaller region than the neutral gas (i.e., \OI\ line). 

\subsubsection{Ancillary Data}
{\it PAH fluxes:} For comparison, we also compile PAH (\citealt{Stierwalt+etal+2014}) and FIR line emission (\citealt{Diaz-Santos+etal+2017}) data from the Great Observatory All-Sky LIRG Survey (GOALS; \citealt{Armus+etal+2009}). Since the Spitzer/IRS spectrum used to measure the PAH emission was extracted from an effective aperture of $\sim 10\arcsec.6\times36\arcsec.6$ \citep{Stierwalt+etal+2014}, and in order to ensure that both the IRS and PACS observations cover the entire source, only sources with $\eta \leq 0.1$ are used in our analysis, where $\eta$ ($\equiv  \log~ F^{\mathrm {IRAC}}_{\mathrm{tot}} [8~\mu {\mathrm m}]/F^{\mathrm{IRS}}_{\mathrm {slit}} [8~\mu {\mathrm m}]$; \citealt{Stierwalt+etal+2014}) is a measure of the total-to-slit flux ratio at 8~$\mu$m. This is because that the PAH-to-IRAC 8 \mum\ flux ratios have a median value of 70\% in normal star-forming galaxies \citep{Smith+etal+2007}, suggesting that $\eta$ is a useful indicator of the fractional power of PAHs covered by the Spitzer/IRS observations. Therefore, we further scale the PAH fluxes by multiplying $10^\eta$. 

{\it Broad-band photometric data:} We compile photometric measurements from far-ultraviolet (FUV) to FIR bands, including the Galaxy Evolution Explorer (GALEX; \citealt{Morrissey+etal+2007}), the Two Micron All Sky Survey (2MASS; \citealt{Skrutskie+etal+2006}), the Hubble Space Telescope (HST), the Spitzer Space Telescope (\citealt{Werner+etal+2004}), the AKARI (\citealt{Murakami+etal+2007}), the Wide-field Infrared Survey Explorer (WISE; \citealt{Wright+etal+2010}), the Infrared Astronomical Satellite (IRAS; \citealt{Neugebauer+etal+1984}), {\it Herschel} and the Submillimetre Common User Bolometer Array (SCUBA) on the James Clerk Maxwell Telescope (JCMT). For  AKARI (\citealt{Kokusho+etal+2017}), IRAS (\citealt{Sanders+etal+2003}) and SCUBA (\citealt{Dunne+etal+2000}) bands, the fluxes are obtained via the NASA/IPAC Extragalactic Database (NED){\footnote{\url{https://ned.ipac.caltech.edu/}}}. For the three HST bands, the fluxes are measured using the images downloaded from the Hubble Legacy Archive website{\footnote{\url{https://hla.stsci.edu/hlaview.html}}}. For the rest bands, the measured fluxes are adopted from the DustPedia project (\citealt{Clark+etal+2018}). All fluxes, which have already been corrected for the Galactic extinction using the extinction law of \cite{Cardelli+etal+1989} assuming RV = 3.1 and adopting $A_V=0.16$~mag from \cite{Schlafly+etal+2011} via the NASA/IPAC Extragalactic Database (NED), and associate uncertainties (including calibration uncertainty) are summarized in Table~\ref{mag-flux}.

\begin{table}
\begin{center}
\caption{NGC~1222 Global Fluxes and Uncertainties}
\label{mag-flux}
  \begin{tabular}{rrrr}
   \hline\noalign{\smallskip}
   \hline\noalign{\smallskip}
\multicolumn{1}{c}{Waveband} & \multicolumn{1}{c}{$\rm \lambda_{eff}$} & \multicolumn{1}{c}{$\rm Flux_{corr}$} & \multicolumn{1}{c}{$\rm Uncertainty_{corr}$} \\
 & \multicolumn{1}{c}{($\rm \mu m$)} & \multicolumn{1}{c}{(Jy)} & \multicolumn{1}{c}{(Jy)}\\
   \hline\noalign{\smallskip}
	GALEX\_FUV &  0.153 & $2.90\times 10^{-3}$ & $1.41\times 10^{-4}$  \\
	GALEX\_NUV &  0.232 & $4.69\times 10^{-3}$ & $1.48\times 10^{-4}$  \\
	HST\_F475W & 0.479 & $2.72\times 10^{-2}$ & $1.11\times 10^{-3}$   \\
	HST\_F606W & 0.596 & $4.20\times 10^{-2}$ & $1.13\times 10^{-3}$   \\
	HST\_F814W & 0.808 & $5.22\times 10^{-2}$ & $1.83\times 10^{-3}$   \\
	2MASS\_J & 1.25 & $8.88\times 10^{-2}$ & $1.74\times 10^{-3}$  \\
	2MASS\_H & 1.65 & $1.12\times 10^{-1}$ & $3.86\times 10^{-3}$  \\
    2MASS\_Ks & 2.17 & $6.69\times 10^{-2}$ & $2.02\times 10^{-2}$  \\
    WISE\_W1 & 3.38 & $5.96\times 10^{-2}$ & $2.23\times 10^{-3}$  \\
    IRAC\_3.6 & 3.56 & $6.21\times 10^{-2}$ & $1.91\times 10^{-3}$  \\
    IRAC\_4.5 & 4.51 & $4.81\times 10^{-2}$ & $1.50\times 10^{-3}$  \\
    WISE\_W2 & 4.63& $4.39\times 10^{-2}$ & $1.86\times 10^{-3}$   \\ 
    AKARI\_S9W & 9.22 & $3.53\times 10^{-1}$ & $1.94\times 10^{-2}$  \\
    IRAS\_12 & 11.69 & $5.00\times 10^{-1}$ & $1.03\times 10^{-1}$ \\
    WISE\_W3 & 12.33& $4.30\times 10^{-1}$ & $2.05\times 10^{-2}$  \\
    AKARI\_L18W & 19.8 & \multicolumn{1}{c}{$1.06$} & $5.48\times 10^{-2}$  \\
    WISE\_W4 & 22.99 & \multicolumn{1}{c}{$1.94$} & $1.09\times 10^{-1}$   \\
    IRAS\_25 & 24.34 & \multicolumn{1}{c}{$2.28$} & $4.58\times 10^{-1}$ \\
    IRAS\_60 & 62.22 & \multicolumn{1}{c}{$13.06$} & \multicolumn{1}{c}{$2.62$} \\
    AKARI\_N60 & 66.73 & \multicolumn{1}{c}{$11.29$} & \multicolumn{1}{c}{$1.61$}  \\
    AKARI\_90 & 89.21 & \multicolumn{1}{c}{$12.61$} & \multicolumn{1}{c}{$1.64$}  \\
    IRAS\_100 & 102.19 & \multicolumn{1}{c}{$15.41$} & \multicolumn{1}{c}{$3.08$} \\
    AKARI\_140 & 149.94 & \multicolumn{1}{c}{$9.49$} & $9.67\times 10^{-1}$ \\
    SPIRE\_250 & 249.36 & \multicolumn{1}{c}{$4.06$} & $2.38\times 10^{-1}$   \\
    SPIRE\_350 & 349.91 & \multicolumn{1}{c}{$1.59$} & $1.06\times 10^{-1}$   \\
    SPIRE\_500 & 504.11 & $5.02\times 10^{-1}$ & $4.84\times 10^{-2}$   \\
    SCUBA\_850 & 850 & $8.40\times 10^{-2}$ & $1.60\times 10^{-2}$  \\
   \noalign{\smallskip}\hline			
\end{tabular}
\end{center}
\end{table}

\section{Results and Analysis}
\label{sect:results}
\subsection{Optical Emission-line Diagnosis}
\label{sect:BPT}
\begin{figure*}[tb]
  \centering
   \includegraphics[width=0.98\textwidth,bb=69 396 557 610]{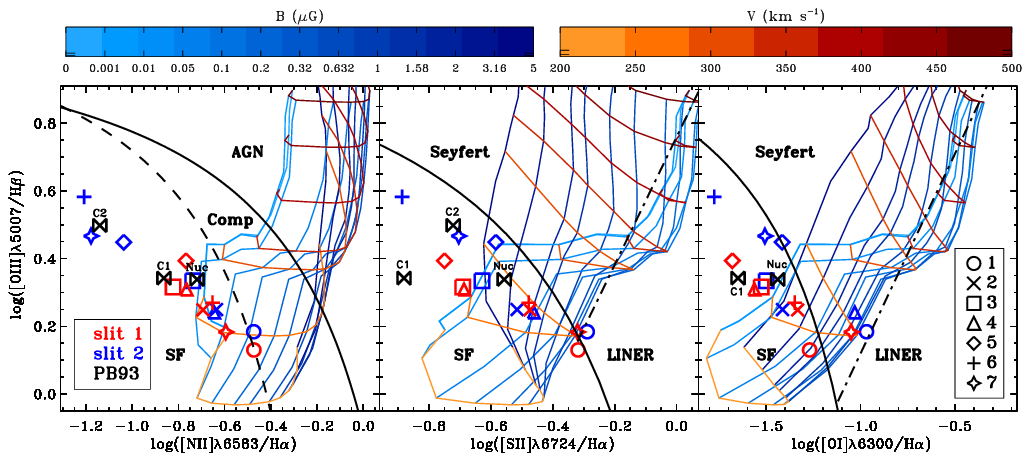}
   \caption{Emission-line diagnostic diagrams for NGC~1222. {\it Left}: Standard BPT diagram. {\it Middle}: [S\,{\sc ii}]/H$\alpha$  versus \OIII/H$\beta$. {\it Right}: [O\,{\sc i}]/H$\alpha$ versus \OIII/H$\beta$. In all three panels, the solid lines separate AGN (above) from star-forming galaxies (below) according to the theoretical models in \citet{Kewley+etal+2001}. The dashed line in the left panel is the division between star-forming and composite systems from \citet{Kauffmann+etal+2003}, and the dash-dotted lines in the right two panels are the theoretical line separating Seyfert galaxies and LINERs from \citet{Kewley+etal+2006}. Different apertures observed with the LJT are demonstrated by the colored points (see the legends in the right panel for details), and the three regions (C1, C2 and nucleus) from \citet{Petrosian+Burenkov+1993} are shown by black bowties. We also overplot the fast shock$+$precursor model grids with solar abundance and preshcok density $n=0.1$ cm$^{-3}$ \citep{Allen+etal+2008} on the \NIIopt/H$\alpha$ (left), \SII/H$\alpha$ (middle) and \OIopt/H$\alpha$ (right) vs. \OIIIopt/H$\beta$ diagrams. The colored lines represent constant shock velocity and magnetic field (as indicated by the color bars on the top), respectively.}
   \label{figbpt}
\end{figure*}

The BPT diagnostic diagram (\citealt{Baldwin+etal+1981}; \citealt{Veilleux+Osterbrock+1987}) is usually used to distinguish the power source of emission lines. Figure~\ref{figbpt} displays \NIIopt/H$\alpha$ versus \OIIIopt/H$\beta$ (left panel), [S\,{\sc ii}]/H$\alpha$ versus \OIIIopt/H$\beta$ (middle panel) and \OIopt/H$\alpha$ versus \OIIIopt/H$\beta$ (right panel) ratios, with red and blue points respectively representing the regions of slits 1 and 2, and black bowties showing the results from \citet{Petrosian+Burenkov+1993}. As defined in \citet{Kewley+etal+2001}, sources lying below and above the solid lines are classified as star-forming (SF) galaxies and AGNs, respectively. The dashed line in the left panel is the empirical line that further divides those SF sources into composite (in between these two lines) and pure SF systems \citep{Kauffmann+etal+2003}, and the dash-dotted lines in the left two panels adopted from \citet{Kewley+etal+2006} is used to separate Seyfert galaxies from LINERs. 

Generally, as shown in Figure~\ref{figbpt}, the optical lines in NGC~1222 are powered by SF, which is consistent with the result in \citet{Riffel+etal+2013} based on the near-infrared diagnostic criteria ([Fe\,{\sc ii}]/Pa$\beta$-H$_2$/Br$\gamma$). Some regions, such as regions 1 and 4 (slit 2; blue circle and triangle, respectively) and 7 (slit 1; red star), however, show ambiguous exciting sources. For example, aperture 7 from slit 1 is located in the SF area on the \NIIopt/\Ha\ plot but in the Seyfert/LINER region on the other two diagrams, whereas aperture 4 from slit 2 is located in the AGN area on the \OIopt/\Ha\ plot but in the SF region on the other two diagrams. This result seems to be consistent with the classification of NGC~1222, a LINER-AGN, in \cite{Nyland+etal+2017}, based on the \OIIIopt/H$\beta$ versus $\sigma_\star$ (stellar velocity dispersion) diagnostic diagram. However, the significantly extended nuclear radio emission with a complex morphology (\citealt{Nyland+etal+2017}) in NGC~1222, which could come from SF, an AGN, or a mixture of these two processes, indicates this LINER-AGN nulcear emission line classification is very uncertain. 

Meanwhile, this contradictory power source might be reconciled with the help of shock heating. Models (\citealt{Allen+etal+2008}) with shock speeds of $v_s = 200-250$ km~s$^{-1}$, preshock density $n=0.1$~cm$^{-3}$ and magnetic fields of $B = 0.0–1~\mu G$ can well reproduce each observed line ratio for regions 4 (slit 2) and 7 (slit 1). To reproduce each observed line ratio for region 1 (circles in Figure \ref{figbpt}) from both slits, it needs a higher preshock density of $n=1$~cm$^{-3}$. The shock-dominated regions are generally located in the outer region of the galaxy, consistent with the results found in \cite{Sharp2010} for several starburst systems via integral field observations. Furthermore, the regions with high \OIopt/\Ha, which is a particularly good tracer of shock excitation (e.g., \citealt{Farage+etal+2010, Rich+etal+2011}), are overlaid with the dust lane, consistent with literature results that shocks are strongly correlated with dust filaments (see, e.g., \citealt{Martini+etal+2003a,Martini+etal+2003b,Farage+etal+2010, Rich+etal+2011, Medling+etal+2021}). Given that shocks are prevalent in the latest-stage mergers (e.g., \citealt{Rich+etal+2015}), it is likely that they also play a role in heating the gas in NGC~1222. Nevertheless, it needs more kinematic information to confirm the existence of shocks, which is hampered by the low resolution of our spectroscopic data. We will investigate the spectral energy distribution (SED), and MIR and FIR line properties to further explore the power source in the following section. 

\subsection{SED Fitting}
\label{SED}
\begin{figure*}[tb]
	\centering
	\includegraphics[bb=90 233 491 528, width=0.75\textwidth]{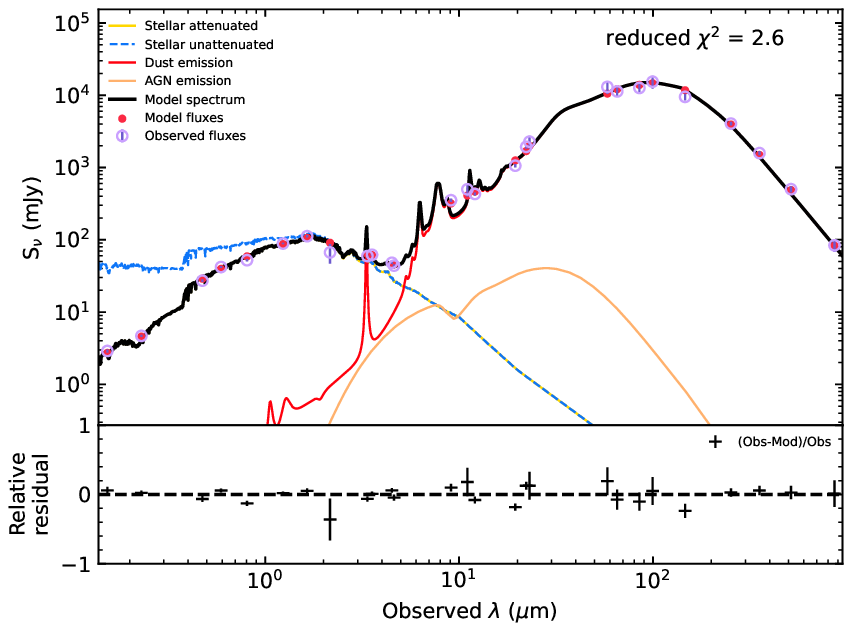}
	\caption{{\it Top:} Panchromatic SED (open circles) for NGC~1222 based on the photometric measurements listed in Tabel~\ref{mag-flux}. The thick black line (and solid circles) is the best-fitting SED model inferred from CIGALE, and the remaining lines are explained in the legend. {\it Bottom:} Relative residuals ($(L_{\mathrm {obs}}-L_{\mathrm {model}})/L_{\mathrm {obs}}$) between the model and the observed fluxes in each wave band.}
	\label{figSED}
\end{figure*}

\begin{table*}
	\begin{center}
		\caption{Derived Properties Parameters From SED Fitting with CIGALE}
		\label{SED-par}
		\begin{tabular}{cccccccc}
			\hline\noalign{\smallskip}
			\multirow{2}{*}{Metallicity}&$M_{\mathrm {dust}}$ & $M_{\star}$ & $L_{\mathrm {dust}}$ & \multirow{2}{*}{$M_{\mathrm {dust}}/M_\star$}& SFR &\multirow{2}{*}{$f_{\mathrm {AGN}}$}&\multirow{2}{*}{$\chi^2_{\mathrm{red}}$} \\
			&$(M_{\odot})$ & $(M_{\odot})$ & $(L_{\odot})$ && $(M_{\odot}~{\mathrm{yr}^{-1}})$ &&\\
			\hline\noalign{\smallskip}
		    \hline\noalign{\smallskip}
			$0.02\pm0.001$&$(1.56\pm0.08)\times10^7$ & $(4.68\pm0.78)\times10^9$ & $(4.15\pm0.21)\times10^{10}$ & $3.3\times10^{-3}$ &$3.72\pm0.95$&$0.01\pm0.002$ & 2.6\\
			\noalign{\smallskip}\hline
		\end{tabular}
	\end{center}
\end{table*}

To further explore the power source(s) in NGC~1222, we construct the FUV-to-submm multi-wavelength SED, and perform fitting with the Python code CIGALE ({\sc version} 2022.1; \citealt{Boquien+etal+2019,Yang+etal+2020}). The models are constructed with the combination of different components, such as SF history (SFH), flexible dust decay curves, dust emission templates, synchrotron emission, AGN contribution, and the intergalactic medium influences (\citealt{Boquien+etal+2019}). Each component is independently computed by different modules. During the fitting process, the {\it sfhdelayed} SFH (\citealt{Boquien+etal+2019}), BC03 (\citealt{Bruzual+Charlot+etal+2003}) SSPs with \citet{Chabrier+etal+2003} initial mass function and two metallicities (0.4 and 1.0 solar abundances), dust emission based on \cite{Draine+Li+2007} library, and \cite{Calzetti+etal+2000} dust attenuation curve, are adopted. Based on the BPT diagram discussed in Section \ref{sect:BPT}, we also add a clumpy AGN component from \cite{Stalevski+etal+2016}. 

In the top panel of Figure \ref{figSED}, the best-fitting SED (black line), the unattenuated stellar population spectrum (blue dashed line), the dust emission (red line) and the AGN component (yellow line) from CIGALE are overlaid on the observed data (open circles). As we can see, the model SED well reproduces the observed fluxes, with the reduced $\chi^2_{\mathrm{red}}=2.6$. The derived parameters from the bes-fit model are listed in Table~\ref{SED-par}. The AGN, if exists, only contributes a very tiny fraction ($f_{\mathrm{AGN}}=0.01\pm0.002$) to the total luminosity. In combination with the aforementioned result based on the BPT diagram, therefore, it is very likely that a shock other than an AGN also contributes to the excitation of the optical lines.

Based on the SED-fitting result, NGC~1222 has a stellar metallicity ($Z_\star=0.02\pm0.001$) around the solar value ($Z_\odot$; $12+\log (\mathrm{O/H})_\odot=8.69$; \citealt{Asplund+etal+2009}) and a stellar mass $M_\star=4.68\times10^9M_\odot$, which agrees with the stellar metallicity-mass ($Z_\star-M_\star$) relation obtained in \cite{Zahid+etal+2017}. The dust mass from the best-fitting is  $M_{\mathrm{dust}}=1.56\times10^7M_\odot$, resulting in a dust-to-stellar mass ratio of $M_{\mathrm{dust}}/M_\star=3.3\times10^{-3}$. This dust-to-stellar mass ratio is comparable to that of late-type galaxies with similar stellar masses (\citealt{Smith+etal+2012}), but is about $40\times$ and $90\times$, respectively, larger than the mean values of the 24 (detected in FIR) and 39 (including sources undetected in FIR) S0 galaxies in  \citet{Smith+etal+2012}. For the gas-to-dust mass ratio (GDR), it is $\sim$320, computed with the aforementioned atomic and molecular gas masses, and is $3\times$ higher than the mean value of the ETG sample in \citet{Smith+etal+2012}. 

We note that, however, the gas-phase abundance ($Z_\mathrm{gas}$) of NGC~1222 is in the range of  $12+\log (\mathrm{O/H})=8.16-8.54$ (i.e., $0.3-0.7Z_\odot$), calculated with the emission-line fluxes from our long-slit spectra and the O3N2 method from \cite{Pettini+Pagel+2004}. However, the metallicities calibrated by this stong-line diagnostics can introduce substantial systematic uncertainties (e.g., \citealt{Kewley+Ellison+2008}). Therefore, we also employ the method proposed in \cite{Dopita+etal+2016} to further check the abundances of various regions, and obtain $12+\log (\mathrm{O/H})=8.0-8.55$, generally consistent with the O3N2 results. Furthermore, 
these metallicities agree with the global measurement of $12+\log(\mathrm{O/H})=8.32\pm0.13$ from \cite{Perez-Diaz+etal+2024}, who derived the chemical abundances using fine-structure lines in the mid-to-far-infrared range. The sub-solar abundance of NGC~1222 suggests that it would have a GDR of $380_{-100}^{+130}$ according to the broken power-law GDR$-$$Z_\mathrm{gas}$ relation given in Table 1 of \cite{Remy-Ruyer+etal+2014}. Therefore, it is very likely that the large difference of GDR between NGC~1222 and the ETG sample in \citet{Smith+etal+2012} is caused by the difference in the metal abundances, as the latter has a mean stellar mass of $\sim$$3\times10^{10}~M_\odot$, resulting in a much higher gas-phase metallicity (super-solar; and thus a lower GDR) according to the $M_\star-Z_\mathrm{gas}$ relation (e.g., \citealt{Tremonti+etal+2004, Zahid+etal+2017}).

It is also interesting to note that the gas-phase abundance is only about $0.5Z_\star$ in NGC~1222, which is opposite to the results presented in \cite{Zahid+etal+2017}, who find that star-forming galaxies generally have $Z_\mathrm{gas}\sim2Z_\star$. This result is very consistent with the scenario that the gas in NGC~1222 was accreted via mergers with gas-rich but metal-poor dwarf galaxies such as C1 and C2. We will further discuss the gas origin in \S\ref{gasflow}.

\begin{figure*}[t]
	\centering
\includegraphics[width=0.95\textwidth,bb=55 342 526  688]{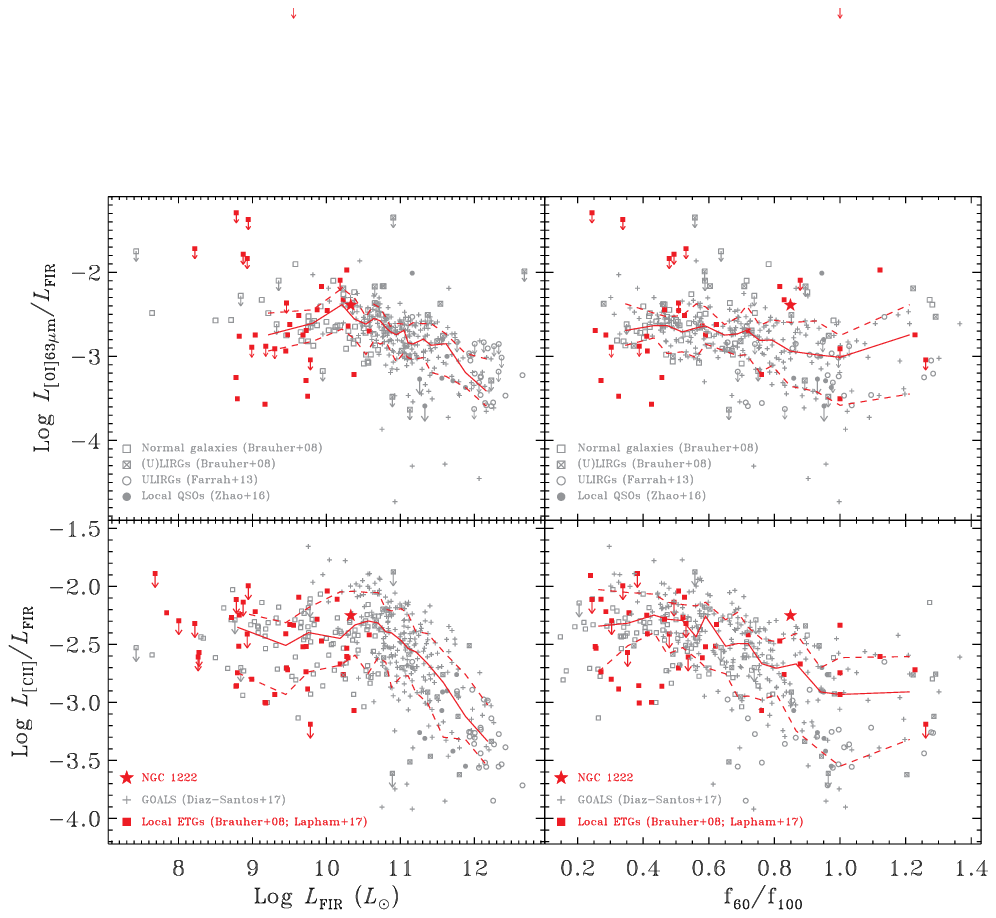}
	\caption{Line-to-FIR ratios vs. \LFIR ({\it left}) and FIR color ($f_{60}/f_{100}$; {\it right}) for various types of galaxies. Red (solid) squares show ETGs from \citet{Brauher+etal+2008} and \citet{Lapham+etal+2017}, open squares are local normal galaxies from \citet{Brauher+etal+2008},  pluses, open circles and crosses display (U)LIRGs from \citet{Diaz-Santos+etal+2017}, \citet{Farrah+etal+2013} and \citet{Brauher+etal+2008}, respectively. Solid circles give QSOs in \citet{Zhao+etal+2016b}. NGC~1222 is shown with the red star. Upper limits are illustrated with arrows. The red solid and dashed lines in each panel respectively illustrate the median and the 16th to 84th percentile range of $y$-axis (without taking account of upper limits).}
	\label{OI_FIR}
\end{figure*}

\subsection{FIR Line-to-continuum Ratios}
As the two main coolants in neutral gas, the combination of the \OI\ and \CII\ lines allows unique determination of the physical properties (e.g. \citealt{Kaufman+etal+1999}) of Photodissociation Regions (PDRs; \citealt{Tielens+Hollenbach+1985}), due to their considerably different excitation temperatures ($\sim$228 and 92~K respectively) and critical densities ($\sim$5$\times10^5$\,cm$^{-3}$ and $\sim$3$\times10^3$\,cm$^{-3}$ respectively; for hydrogen nuclei). Furthermore, the ratio of the total luminosity of these two lines to the IR continuum ($(L_{[\mathrm{C}\,\scriptsize{\textsc{ii}}]}+L_{[\mathrm{O}\,\scriptsize{\textsc{i}]\,63\,\mu{\mathrm{m}}}})/L_{\mathrm{IR}}$) can be used as a measure of the PE heating efficiency (e.g. \citealt{Tielens+Hollenbach+1985}), which represents the far-ultraviolet (FUV) energy that goes into to the gas, divided by the FUV energy deposited in dust grains.

In Figure~\ref{OI_FIR}, we show the \OI\ (upper) and \CII\  (bottom) to FIR luminosity ratios as a function of \LFIR\ ($42.5-122.5$~\mum; left) and $f_{60}/f_{100}$ (right) for various types of local galaxies, i.e., ETGs (\citealt{Brauher+etal+2008}; \citealt{Lapham+etal+2017}), normal galaxies (total infrared luminosity $L_{\mathrm{IR}}<10^{11}~L_\odot$, hereafter excluding ETGs unless stated explicitly; \citealt{Brauher+etal+2008}), (Ultra-)luminous infrared galaxies ((U)LIRGs; \citealt{Brauher+etal+2008};  \citealt{Farrah+etal+2013}; \citealt{Diaz-Santos+etal+2017}) and infrared luminous QSOs \citep{Zhao+etal+2016b}. In each panel, the red solid line plots the median value, and the dashed lines give the 16th to 84th percentile range. From the left two panels of Figure~\ref{OI_FIR} we can see that the line-to-FIR luminosity ratios show a large variation at a given luminosity. Compared with late-type galaxies, in general, ETGs have similar \LCII/\LFIR\ but lower \LOI/\LFIR\ ratios, and display a wider spread of \LOI/\LFIR. However, NGC~1222 is one of a few ETGs 
having the largest \LOI/\LFIR\ and \LCII/\LFIR\ ratios, with the line-to-FIR ratios about 0.3 dex larger than the median values of all ETGs. NGC~1222 also has its line-to-FIR ratios $\sim$2 times larger than the median values for those sources with $|\log (L_{\mathrm{FIR}}/L_{\mathrm{FIR,NGC1222}})|\leq0.3$. 

The higher-than-average line-to-FIR ratios of NGC~1222 is more prominent in the right panels of Figure \ref{OI_FIR}. Among the sources with a similar $f_{60}/f_{100}$ ($\sim$0.85), NGC~1222 has the largest \LCII/\LFIR\ ratio, which is about 0.5 dex higher than the median value. There are only a handful objects, including NGC~1222, also showing such a large deviation from the median curve of \LOI/\LFIR\ at this FIR color. Therefore, these results suggest that the FIR line emissions in NGC~1222 seem to be enhanced, compared with the galaxies at similar redshifts with similar \LFIR\ and ``effective" FIR dust temperature (i.e., FIR $f_{60}/f_{100}$ colors). It is worth noting that among the eight normal galaxies (including 3 ETGs here but exclusive of NGC~1222) with $\gtrsim$0.4 dex higher-than-median \LOI/\LFIR\ at corresponding $f_{60}/f_{100}$, seven are found evidences of shock heating in the literature works (i.e., \citealt{Omar+etal+2002}: Mrk~1; \citealt{Bogdan+etal+2017}: Mrk~3; \citealt{Mingo+etal+2011}: Mrk~6; \citealt{Paggi+etal+2012}: Mrk~573; \citealt{Mundell1999} and \citealt{Wang+etal+2011}: NGC~4151; \citealt{Croston+etal+2008}: NGC~6764; \citealt{Mortazavi2019}: NGC~7714). The remaining source, NGC~2415, is a compact, interacting starburst galaxy hosting core-collapse supernovae (SN; \citealt{Kangas+etal+2013} and references therein),  and thus stellar wind- and SN-driven shocks could be expected (see \citealt{Thompson2024} for a review). 

\subsection{Shock-Enhanced Line Emissions in NGC~1222?}
\label{shock_ir}
\begin{figure}[t]
	\centering
	\includegraphics[bb=46 382 459 765, width=0.47\textwidth]{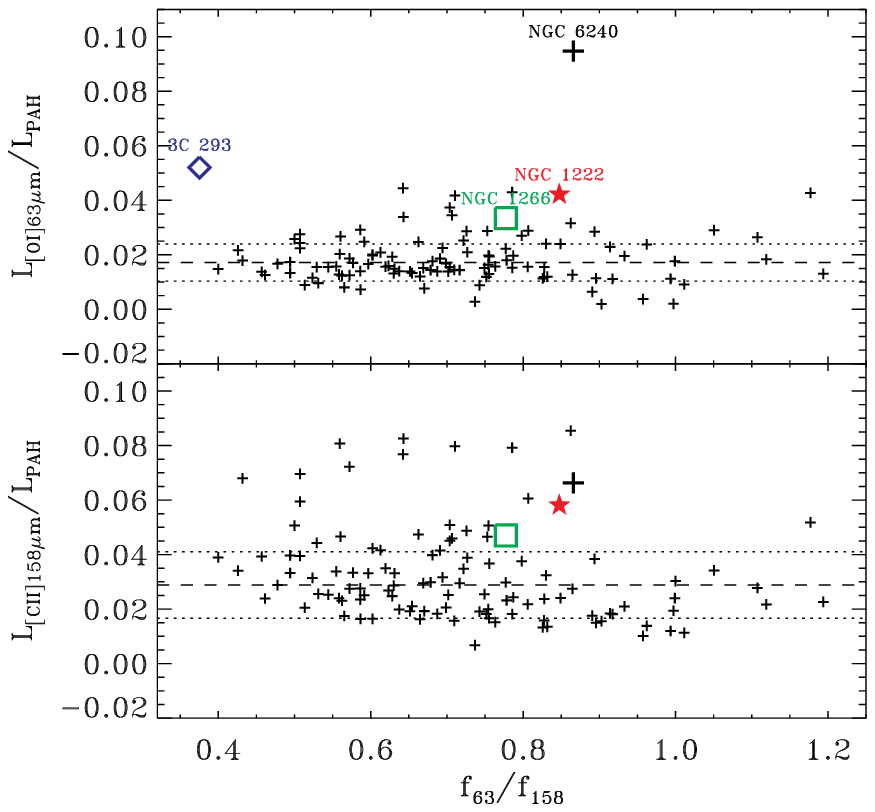}
	\caption{The \OI\ (upper) and \CII\ (bottom) line-to-PAH luminosity ratios plotted as a function of the FIR color $f_{63}/f_{158}$ for GOALS LIRGs (plus signs). The dashed and dotted lines represent the mean value and $\pm1\sigma$ deviation respectively. The vertical spans of both plots are the same for direct comparison. The three shock-excited sources (diamond: 3C~293, square: NGC~1266 and bigger plus: NGC~6240) are also marked with their corresponding names.}
	\label{line2pah}
\end{figure}

The \OI\ and \CII\ lines are thought to be excited by collisions with atomic and molecular hydrogen in PDRs, where the gas are heated by the PE effect, i.e., hot electrons are ejected from PAH molecules and small dust grains by absorbing FUV photos, and transfer their energy to the gas. It is also shown theoretically that these two lines can be enhanced by low-velocity ($v_s = 3-40$ \kms; e.g., \citealt{Lesaffre+etal+2013}) and fast ($v_s = 30-150$ \kms; e.g., \citealt{Hollenbach+McKee+1989}) shocks, as observed in some merging/interacting systems (e.g., \citealt{Appleton+etal+2013, Fadda+etal+2023}). 

Specifically for NGC~1222, we have demonstrated that it seems to need a shock to help explain the observed optical line ratios in some regions, and the FIR line-to-continuum ratios are also $2-5\times$ boosted compared with the median values of the galaxies with similar \LFIR\ or $f_{60}/f_{100}$. In addition, observations (\citealt{Appleton+etal+2013, Fadda+etal+2023}) have shown that \OI/PAH and/or \CII/PAH ratios for shocked gas may be elevated, about $3-20$ times higher than those produced by PE heating in PDRs. In Figure \ref{line2pah}, we plot the line-to-PAH ratios as a function of the FIR 63-to-158 \mum\ color ($f_{63}/f_{158}$) for the GOALS LIRGs sample (\citealt{Diaz-Santos+etal+2017}) and two shock-excited galaxies: 3C~293, a radio galaxy possibly having radio jet-driven shocks (\citealt{Ogle+etal+2010}), and NGC~1266, an early-type galaxy excited by outflow-driven shocks (\citealt{Pellegrini+etal+2013, Chen+etal+2023}). We also mark NGC~6240, in which the starburst superwinds power the large-scale diffuse ionized gas (e.g., \citealt{Heckman+etal+1987}) and shock-excited molecular gas emission (e.g., \citealt{Max+etal+2005, Meijerink+etal+2013}). 

As shown in the upper panel of Figure \ref{line2pah}, we can see that the \OI/PAH ratios of the three shock-excited sources are about 2 (NGC~1266) to 5.5 (NGC~6240) times higher than the mean value (0.017) of the GOALS sample. Here we have computed the sample mean and standard deviation with the robust mean and 3$\sigma$ clipping method, due to the fact that shocks are also present in some (U)LIRGs (\cite{Rich+etal+2015}). Similar to the above three sources, the observed \OI/PAH ratio (0.042) for NGC~1222 is about 2.5$\times$ ($\sim$3.6$\sigma$ with $\sigma=0.007$) higher than the mean ratio, suggesting that shocks may also play a role in exciting the gas emission.

Regarding the \CII\ emission, as demonstrated in the bottom panel of Figure \ref{line2pah}, it has a relatively larger spread. Furthermore, the shock-enhancement ($1.6\times$ and $2.3\times$ for NGC~1266 and NGC~6240 respectively) is smaller compared with the \OI\ line. For NGC~1222, the \CII/PAH ratio is in between those of NGC~1266 and NGC~6240, and is about 2 times higher than the mean value. These results agree with the shock models computed numerically in \cite{Lesaffre+etal+2013}, who have shown that the \OI\ emission is generally enhanced by a larger amplitude than the \CII\ line. Therefore, it is likely that shocks, which may arise from the merging process, play a role in exciting both the optical lines and molecular gas in NGC~1222.

\subsection{Evidence for Gas Inflow}
\label{gasflow}
\begin{figure}[tb!]
\centerline{\includegraphics[width=0.47\textwidth,bb=37 368 560 922]{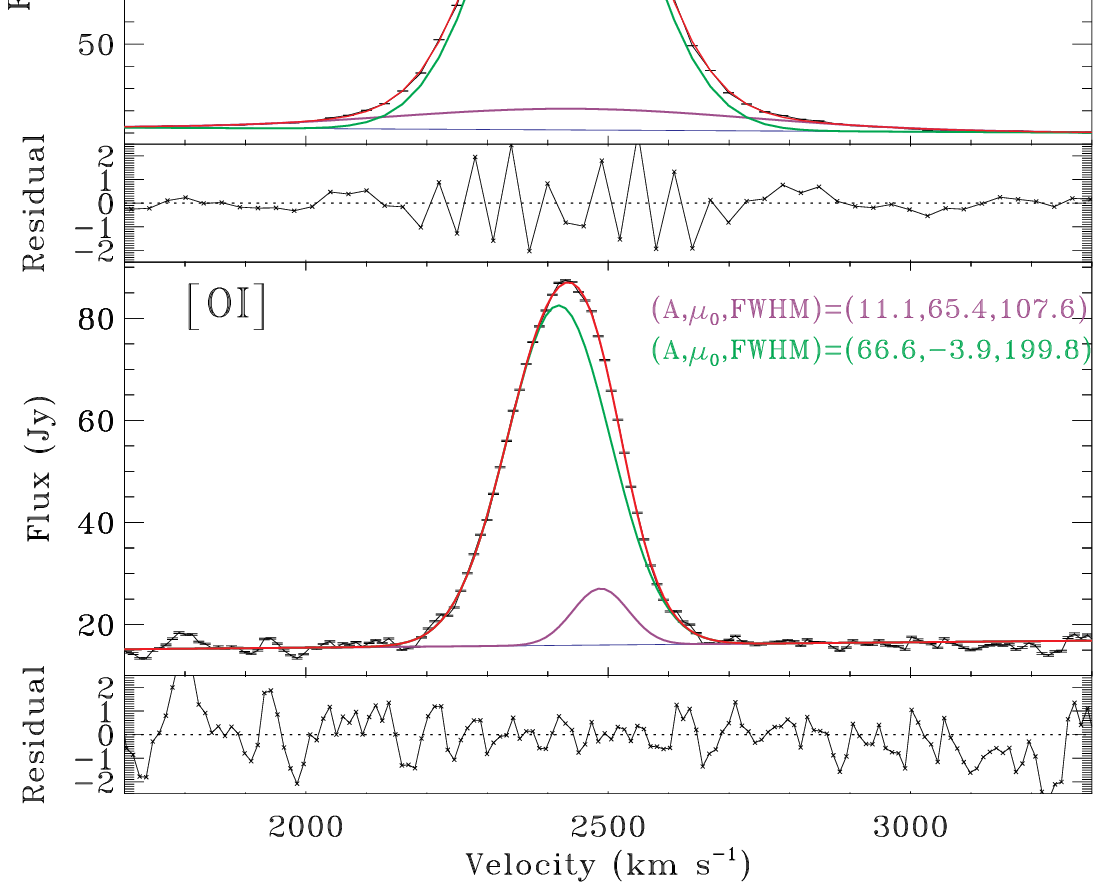}}
\caption{Integrate PACS spectra of \OI\ (bottom) and \CII\ (upper) for NGC~1222. Overlaid are the best fitted results with two Gaussian functions for the line emission. The best-fit parameters are also labelled, where the central velocities ($\mu_0$) are relative to the systematic velocity of 2422~\kms, and the FWHMs have not been corrected for the instrumental spectral resolutions (about 88 \kms\ and 238 \kms\ respectively for \OI\ and \CII).}\label{pacs_spec}
\end{figure}

Using high resolution ($\sim$3\arcsec.5) interferometric data, \cite{Alatalo+etal+2013} find that NGC~1222 shows highly disrupted CO distribution and kinematics (e.g., see Figure \ref{pacs_lines}d). Combining this with the fact that NGC~1222 is an interacting system confirmed by optical observations (e.g., \citealt{Petrosian+Burenkov+1993, Duc+etal+2015}), the authors suggest that the gas in NGC~1222 was acquired via external accretion, which is also evidenced by the 4$\times$ lower-than-average gas-phase to stellar abundance ratio discussed in \S\ref{SED}. To further investigate whether there is any sign of gas inflow, first we check the integrated spectra of \OI\ and \CII\ lines, as shown in Figure \ref{pacs_spec}.

From the figure we see that two Gaussian components can well fit the line profile for both lines, although the residual spectrum of \CII\ shows a larger scatter than the line-free window (top two panels of Figure \ref{pacs_spec}). For the main component, the centers of both lines agree well with each other, and are very close to the mean stellar velocity of 2422 \kms\ (\citealt{Cappellari+etal+2011}). For the second component, however, the \CII\ line is much broader but has a smaller red-shifted velocity than \OI\ (bottom two panels). The flux ratios of the second-to-main component are 0.15 and 0.09 for \CII\ and \OI\ respectively. These results are consistent with a gas inflow scenario, in the sense that the \OI\ line is optically thick and thus the observed flux mainly comes from the front side of the galaxy, while the \CII\ line is thought to be marginally optically thick (\citealt{Kaufman+etal+1999, Luhman+etal+2003}) and hence the observed flux can include emission from both sides.

\begin{figure*}[tb!]
\centerline{\includegraphics[width=0.85\textwidth,bb=28 253 515 511]{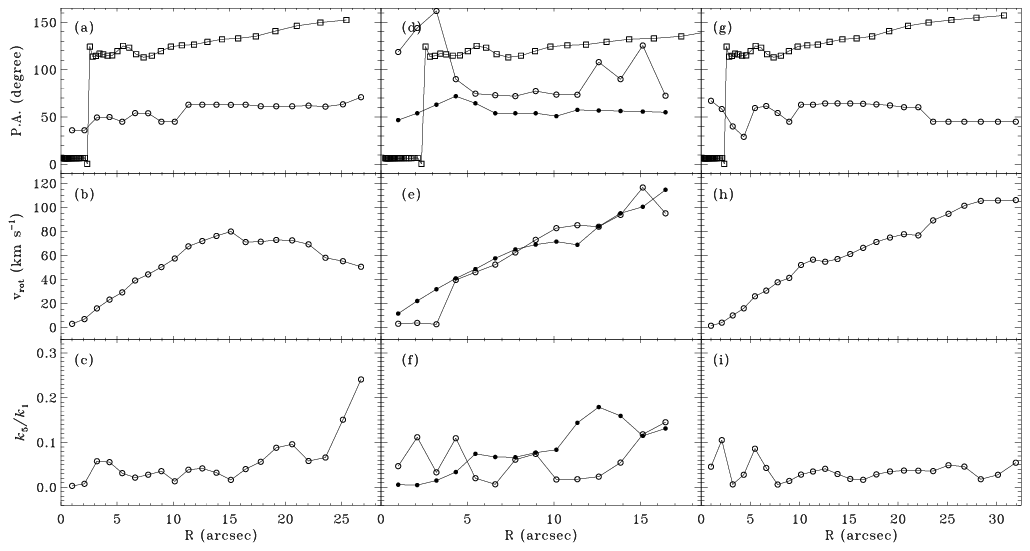}}
\caption{Kinemetric coefficients of the \OI\ ({\it left}), \OIIIopt\ (solid circles) and \NI\ (open circles; {\it middle}), and \CII\ ({\it right}) velocity maps. (a), (d) and (g): Kinematic P.A.; (b), (e) and (h): projected rotation speed; and (c), (f) and (i): magnitude of non-circular motions, i.e., the deviation from simple rotation. For comparison, the position angles obtained from the best-fitting photometric ellipses ($r$-band) are also overlaid (squares in the top panels).}\label{fig_kinemetry}
\end{figure*}

The external gas origin can be further checked by comparing the kinemetric features between the gaseous and stellar components. To this end, we use the code {\it kinemetry},
which does surface photometry to the first moment map by determining the best-fitting ellipse, to extract information such as mean velocity ($v_{\mathrm {sys}}$), rotation curve ($v_{\mathrm{rot}}$), kinematic position angle (P.A.; north to east), deviation from simple rotation ($k_5/k_1$), etc (see \citealt{Krajnoic+etal+2006} for more details). The kinemetric results for the gas emission, i.e., \OI, \OIIIopt\ (\citealt{Cappellari+etal+2011}) and \NI, and \CII\ velocity maps, are shown in the left, middle and right column of Figure \ref{fig_kinemetry}, respectively. 

The top row of Figure \ref{fig_kinemetry} presents the kinematic P.A. (hereafter PA$_{\mathrm{kin,gas}}$) as a function of radius $R$, where the center was set to the optical center. We can see that PA$_{\mathrm{kin,gas}}$ for all but the \NI\ lines only show small variations and are consistent with each other, with median values of $61^\circ\pm3^\circ$, $55^\circ\pm3^\circ$, and $60^\circ\pm7^\circ$ for the \OI, \OIIIopt\ and \CII\ lines, respectively. Whereas for the \NI\ velocity map, its P.A. has a larger amplitude of variation (median value of $83^\circ\pm17^\circ$), which may be caused by its relatively poor S/N. 

However, the optical P.A. (hereafter PA$_{\mathrm{phot}}$), which are measured from the best-fitting $r$-band photometric ellipses using {\it AutoProf} (\citealt{Stone+etal+2021}) and generally agree with the kinemetric P.A. (hereafter PA$_{\mathrm{kin,stellar}}$) of the stellar velocity map (e.g., \citealt{Krajnoic+etal+2011}), differ by $\sim70^\circ-100^\circ$ from PA$_{\mathrm{kin,gas}}$. For NGC~1222, it is a featureless non-regular rotator and the stellar rotation is difficult to detect (\citealt{Krajnoic+etal+2011}), whereas both the neutral and ionized gas components show regular rotation (see Figures \ref{fig_kinemetry}(b), (e) and (h)). Combining this with the misalignment of the gas and stellar components (assuming $\mathrm{PA}_{\mathrm{kin,stellar}} = \mathrm{PA}_{\mathrm{phot}}$), it further supports the external gas origin for NGC~1222. 

In the bottom panels of Figure \ref{fig_kinemetry}, we also plot the magnitude of the higher-order term $k_5$ that describes the deviation of isophote shape from an ellipse. As showing in \cite{Krajnoic+etal+2006}, $k_5$ is sensitive to the existence of separate kinematic components on the velocity map. From Figures \ref{fig_kinemetry}(c), (f) and (i) we can see that $k_5$ can be as large as $\sim$25\% of the rotating velocity ($\equiv k_1$) for \OI\ line, which indicates that there may exist some non-circular motions such as infollows in the gas disk, consistent with the accretion scenario. Therefore, it is very likely that the shocks in NGC~1222 are induced by gas inflows.

\section{Discussion}
\label{sect:discuss}
As explained in the previous section, it seems to need additional heating sources, such as shocks, other than the PE process to excite the optical and FIR line emissions in NGC~1222. However, the old stellar populations (\citealt{Cid+etal+2011}), or low luminous AGNs may also be used to interpret the observed optical line ratios. Furthermore, the contribution of the diffuse ionised gas (DIG; e.g., \citealt{Zhang+etal+2017}) and of the PDRs (e.g., \citealt{Sorzer+etal+2000}) can increase the \SII/\Ha\ and \OIopt/\Ha\ ratios. In this section, therefore, we first discuss the contribution of these potential candidates to the optical line ratios. Then we discuss why shock is the dominant source of the infrared line emission, but does not dominate the excitation of the optical lines. 

\cite{Yan+etal+2012} and \cite{Zhang+etal+2017} have shown that old stellar populations and DIG will produce LINER/AGN-like \NIIopt-, \SII- and \OIopt-to-\Ha\ ratios simultaneously. For NGC~1222, however, these three line ratios of several apertures are located in ambiguous regions, ruling out that the old stellar populations and/or DIG dominate the line excitation. This argument can be further confirmed by the fractional contribution of DIG (5.6\%; \citealt{Lapham+etal+2017}) to the \CII\ line, and by the fact that the \Ha\ equivalent width from the shocked regions is much higher than that ($\lesssim$3 \AA; \citealt{Cid+etal+2011}) produced by old stellar populations. Combining the ambiguous powering sources from the BPT diagrams with $f_{\mathrm{AGN}}=0.01$, it is also unlikely that an AGN is present in NGC~1222. 

Regarding the contribution from PDRs, \cite{Sorzer+etal+2000} have modeled the \OIopt\ and \SII\ line emission for a wide parameter range (i.e.,  the total hydrogen density $10^4 \leq n~(\mathrm{cm}^{-3}) \leq 10^7$ and the strength of the FUV field $G_0$ in terms of the Habing field $10^3<G_0<10^7$). The authors find that \OIopt/\OI\ intensity ratio is in the range of $0.001-1.0$ when $n \geq 10^5$ cm$^{-3}$ and $G_0>10^{3.5}$. For NGC~1222, the \OI\ line has a flux of $f_{\mathrm{[OI]63}}=2.52\times10^{-15}$~W~m$^{-2}$, corresponding to $f_{\mathrm{[OI]6300,PDR}}=2.52\times10^{-15}$~erg~s$^{-1}$~cm$^{-2}$ by adopting \OIopt/\OI=0.001 (for $n\sim10^5$~cm$^{-3}$ and $G_0>10^{3.5}$). However, this $f_{\mathrm{[OI]6300,PDR}}$, which is $>$10 times smaller than that ($2.6\times10^{-14}$~erg~s$^{-1}$~cm$^{-2}$) from slit 2, should be considered as an upper limit since the PDR modeling results from \cite{Lapham+Young+etal+2019} give $G_0 \leq 10^{2.75}$ and $n\leq10^{4.25}$~cm$^{-3}$ based on different PDR diagnostics. Therefore, the overall PDR contribution to \OIopt\ is negligible in NGC~1222.

Could aperture 4 of slit 2 be a different case, i.e., having a much higher $n$? To answer this question, we exploit PDRT\footnote{PDRT Toolbox website: \url{https://dustem.astro.umd.edu}} to extract $n$ and $G_0$ for different regions following \cite{Zhao+etal+2016a}. Before modeling, the resolutions of FIR images are matched using the method presented in \cite{Aniano+etal+2011}. We find $n\sim(1-8)\times10^3$~cm$^{-3}$ and $G_0\sim(2-4.5)\times10^2$, hence PDRs only contribute $\lesssim$1.6\% to the total \OIopt\ flux within aperture 4 of slit 2, according to the \cite{Sorzer+etal+2000}'s models and the observed \OIopt\ ($4.3\times10^{-15}$~erg~s$^{-1}$~cm$^{-2}$) 
and \OI\ ($6.8\times10^{-14}$~erg~s$^{-1}$~cm$^{-2}$) 
line fluxes.  

Though shock may play a role in heating both the optical and infrared emission lines in NGC~1222, it seems much less important to excite the optical lines than the infrared line emission as demonstrated in \S\ref{sect:BPT} and \S\ref{shock_ir}. This looks contradictory at the first sight. Firstly, however, our long-slit spectroscopic observations only covers a small fraction of the whole galaxy, and thus it is highly likely that many shocked regions are not identified. Secondly, it needs fast shock ($\gtrsim 100-200$~\kms; e.g., \citealt{Allen+etal+2008}) to significantly enhance the \OIopt\ line emission, which might be rare in NGC~1222 given that the full width at half maximum of the optical lines is only about 100~\kms\ (\citealt{Petrosian+Burenkov+1993}). In contrast, the FIR line emission can be significantly enhanced by low-velocity shocks ($3-40$~\kms; e.g., \citealt{Lesaffre+etal+2013}), which seem to be prevalent in NGC~1222 as evidenced by the non-circular motions illustrated in Figures \ref{fig_kinemetry}(c) and (i). The amplitude of these motions is in the range of several to a few tens \kms, agreeing well with the low-velocity shock scenario. Nevertheless, it needs high-resolution, integral field observations to further explore the detailed shock excitation in NGC~1222.

\section{Summary}
\label{sect:summ}
In this paper we performed a comprehensive study on the ISM properties of NGC~1222, a merging galaxy that forms a triple system with two dwarf components, using optical and FIR spectra, as well as multi-band photometric data. We obtain the dust luminosity, stellar and dust mass, and GDR in NGC~1222 by fitting the observed SED. We discuss gas heating sources via the BPT diagram by using the observed optical emission line ratios, and by comparing the FIR line-to-continuum and line-to-PAH ratios with other galaxies. We also extract the kinematic properties of the gas emission. Our main results are:
\begin{enumerate}
    \item Comparing the optical emission-line ratio with shock models, we suggest that a merger-induced shock is a probable cause of heating the gas in some regions of NGC~1222.
    \item NGC~1222 has a dust-to-stellar mass ratio $M_\mathrm{dust}/M_\star \sim 3.3\times10^{-3}$, which is comparable to late-type sources with similar stellar mass but about 40 to 90 times larger than the mean value of local S0 galaxies. The gas-to-dust mass ratio is about 320, agreeing with that of late-type galaxies with similar metal abundance, and indicating a large gas reservoir.
    \item The stellar metallicity obtained via SED-fitting is nearly the Solar value and consistent with the $M_\star-Z_\star$ relation, but the gas-phase abundances vary between $0.3-0.7Z_\odot$ and are much lower than that expected from the $M_\star-Z_{\mathrm{gas}}$ relation, indicating an external gas origin.
    \item The \OI- and \CII-to-FIR continuum ratios of NGC~1222 is about 2$-$3 times higher than the median value of those galaxies with similar IR luminosity or similar FIR color. NGC~1222 is also among a few galaxies having the highest line-to-PAH ratios, in which shocks are usually present.
    \item There are two components in the \OI\ and \CII\ integrated line profiles. Both the ionized and neutral gas emissions show rotating velocity fields with considerable non-circular motions. There also exists a kinematic misalignment between the gas and stellar components. These results further support gas inflows in NGC~1222. 
\end{enumerate}

\begin{acknowledgments}
We thank the anonymous referee for useful comments/suggestions that significantly improved the paper. This work is supported by the National Natural Science Foundation of China (NSFC grand No. 12173079). We acknowledge the science research grants from the China Manned Space Project with Nos. CMS-CSST-2021-A06. This research use the NASA/ IPAC Extragalactic Database (NED), which is operated by the Jet Propulsion Laboratory, California Institute of Technology, under contract with the National Aeronautics and Space Administration. PACS has been developed by a consortium of institutes led by MPE (Germany) and including UVIE (Austria); KU Leuven, CSL, IMEC (Belgium); CEA, LAM (France); MPIA (Germany); INAF-IFSI/OAA/OAP/OAT, LENS, SISSA (Italy); and IAC (Spain). This development has been supported by the funding agencies BMVIT (Austria), ESA-PRODEX (Belgium), CEA/CNES (France), DLR (Germany), ASI/INAF (Italy), and CICYT/MCYT (Spain). SPIRE has been developed by a consortium of institutes led by Cardiff University (UK) and including Univ. Lethbridge (Canada); NAOC (China); CEA, LAM (France); IFSI, Univ. Padua (Italy); IAC (Spain); Stockholm Observatory (Sweden); Imperial College London, RAL, UCL-MSSL, UKATC, Univ. Sussex (UK); and Caltech, JPL, NHSC, Univ. Colorado (USA). This development has been supported by national funding agencies: CSA (Canada); NAOC (China); CEA, CNES, CNRS (France); ASI (Italy); MCINN (Spain); SNSB (Sweden); STFC and UKSA (UK); and NASA (USA). HIPE is a joint development by the Herschel Science Ground Segment Consortium, consisting of ESA, the NASA Herschel Science Center, and the HIFI, PACS, and SPIRE consortia. Based on observations collected at the Centro Astronómico Hispano Alemán (CAHA) at Calar Alto, operated jointly by the Max-Planck Institut für Astronomie and the Instituto de Astrofísica de Andalucía (CSIC). The Legacy Surveys imaging of the DESI footprint is supported by the Director, Office of Science, Office of High Energy Physics of the U.S. Department of Energy under Contract No. DE-AC02-05CH1123, by the National Energy Research Scientific Computing Center, a DOE Office of Science User Facility under the same contract, and by the U.S. National Science Foundation, Division of Astronomical Sciences under Contract No.AST-0950945 to NOAO.
\end{acknowledgments}

\end{document}